# A constructive formulation of the one-way speed of light

Chandru Iyer
*Techink Industries, C-42, Phase-II, Noida, India 201305*
*Email:* chandru_i@yahoo.com

G. M. Prabhu
*Department of Computer Science, Iowa State University, Ames, Iowa 50011*
*Email:* prabhu@cs.iastate.edu

A formulation of the one-way speed of light in three-dimensional Euclidean space is derived by a constructive approach. This formulation is consistent with the result of the Michelson-Morley experiment in that the harmonic mean of the outward and return speeds is equal to $c$, the standard value for the speed of electromagnetic radiation in vacuum. It is also shown that a shift in synchronization, proportional to the distance along the line of motion, renders this speed a constant along all directions.

## I. INTRODUCTION

It is well known that the Michelson-Morley experiment implies that the average round-trip speed of light is a constant in any inertial frame.[1] As a result of this experiment and the understanding of its implications, there has been much discussion about the one-way speed of light.[2–8] That the speed of light is a constant and the same constant in every direction in every inertial frame is a postulate of special relativity. According to this postulate, light rays, originating from a light source that is at rest in a particular inertial frame, propagate at the same speed in all inertial frames. During the discussions leading to this postulate, Einstein[9] stated "Who would imagine that this simple law has plunged the conscientiously thoughtful physicist into the greatest intellectual difficulties?" These difficulties are related to the problem of the synchronization of spatially separated clocks.[2,10,8] To measure the speed of a moving object, an inertial frame requires a set of spatially separated and synchronized clocks. Any inaccuracy in the synchronization of these spatially separated clocks alters the observed speed of moving objects. Thus, the observed speed of light is related to the methodology adopted in synchronizing spatially separated clocks. The "intellectual difficulties of the physicist"[9] regarding the speed of light and the "persistent controversy"[2] regarding clock synchronization are closely related.

Einstein resolved this problem by specifying an operational procedure to synchronize spatially separated clocks in an inertial frame, a procedure referred to as Einstein synchronization. The procedure is to adjust clock 2 so that the time for light to propagate from clock 1 to clock 2, $t_{out}$, is the same as the time for light to propagate from clock 2 to clock 1, $t_{in}$. Einstein referred to this procedure by the German word "*festsetzung*," which has been translated as "convention," "stipulation," and "definition."[4] The term convention has been widely used, and Janis[10] has discussed the ongoing debate in detail. The Einstein synchronization procedure, formulated independently, is consistent with the conclusion from the Michelson-Morley experiment[1] that the average round-trip speed of light in any closed path in any inertial frame is the same constant $c$. The procedure defines clock synchronization along any small incremental straight line element of a closed path in such a way that the speed of light along the line element is $c$. In other words, it defines the speed of light in all directions to be the same constant in all inertial frames, and utilizes this definition to synchronize spatially separated clocks in any inertial frame. According to this definition the average round-trip speed of light is the same constant in any closed path in all inertial frames, consistent with the Michelson-Morley experiment. However, it imposes the additional constraint that the speed of light is the same constant in any direction in any inertial frame. This additional constraint is not a direct consequence of the Michelson-Morley experiment, but a consequence of the implicit expectation that all directions are equivalent in any inertial frame.

Reichenbach[7] argued that Einstein's procedure involves circular reasoning and proposed a different synchronization convention involving a parameter ε in the range $0 < ε < 1$. Reichenbach synchronization assumes that the outward and return speeds of light are different. The discussions are usually restricted to one-dimensional space. In the context of dynamics, Ohanian[2] has argued that the Einstein synchronization convention, which forces the one-way speed of light to be the same constant in all inertial frames, must be used to obtain the standard form for Newton's second law of motion. However, Martinez[4] and Macdonald[3] commented that dynamical considerations cannot be more fundamental than kinematical considerations in defining a synchronization convention. Ohanian[5] replied that a preferred synchronization is akin to a preferred coordinate system and "some choices of coordinates and of synchronization play a preferential role, because they permit us to express the laws of physics in their simplest form."

There are two methods to achieve synchronization of spatially separated clocks in an isotropic inertial frame. An isotropic inertial frame is defined as an inertial frame in which the choice of a direction does not influence the outcome of any physical process. One method is by the slow separation of clocks. The other is by synchronizing clocks assuming that the speed of light is constant and independent of direction. Even though both these methods achieve identical synchronization in a given inertial frame, some believe that both can be in error by the same amount in an anisotropic inertial frame.[11] Eddington[6, 10] stated that "they (both) are best regarded as conventional." Bohm[12] has given a description, based on the Lorentz[13, 14] theory of electrons, of the process of synchronization by slow separation of clocks and has shown that the synchronization achieved by slow separation of clocks is different for two inertial frames in relative motion. Bohm[12] pointed out that every inertial frame that synchronizes spatially separated clocks by assuming it is isotropic will find that all other inertial frames moving relative to it are anisotropic. An inertial frame will find itself to be anisotropic only when it accepts the intrinsic isotropic clock synchronization of another inertial frame moving relative to it. In summary, all inertial frames are intrinsically isotropic, and each one of them finds all other inertial frames with relative motion to be anisotropic. One of the manifestations of this anisotropy is that the speed of light is different in different directions. There can be other manifestations such as the directional dependence of dynamical phenomena in an anisotropic reference frame.[2]

## II. SYNCHRONIZATION BY THE METHOD OF SLOW SEPARATION OF CLOCKS

We consider motion along a line, the *x*-axis, in an inertial frame. If two inertial frames in relative motion are involved, we restrict our discussions to the line of relative motion. We revisit Bohm's[12] process of generating a set of spatially separated and synchronized clocks by the procedure of slow separation of clocks and emphasize that it is successful only in an isotropic inertial frame.

It is well established that moving clocks are observed to run slow. An inertial frame can be specified by a set of observers and objects that are stationary with respect to each other. Any clock that moves with respect to this set of stationary observers is considered to be moving with respect to that inertial frame. Time dilation may be defined as the ratio of the time elapsed between two events occurring at the location of the moving clock according to an observer co-moving with the moving clock to the time elapsed between those two events as observed by stationary observers. A clock that has a velocity of v in an isotropic inertial frame K can only have a time dilation that is a positive and even function of v, because any component that is an odd function of v will contribute to anisotropy. The time dilation function could contain a first-order term in $|v|$ and still be isotropic. However, this possibility renders the time dilation function continuous but not differentiable at v = 0. We exclude this possibility[15] and consider the time dilation function to be a positive and even function of v. This function can be expanded as a Taylor series in positive even powers of v. When we consider a clock moving at a small velocity Δv, the (proper) time for a given separation s is (s/Δv) as measured by clocks in a stationary inertial frame. The time difference between that measured by the moving clock and the proper time of the stationary inertial frame is (s/Δv) multiplied by a series of positive even powers of Δv. After cancellation of Δv in the denominator, the expression reduces to a series of positive odd powers of Δv. This series approaches to zero as Δv goes to zero. Thus one can separate clocks slowly enough to preserve their synchronization in an isotropic inertial reference frame. The desired accuracy determines how slow the separation should be. In the limit, when the speed of separation tends to zero, the inaccuracy in the synchronization also tends to zero. The principle may be stated as follows: In an isotropic inertial frame, when identical and synchronized clocks initially at the same location are separated very slowly, the synchronization of these clocks is almost unaffected and the error inherent in this procedure is of order sΔv/$c^2$, where s is the distance of separation.

When the inertial frame is anisotropic, the time dilation function contains both odd and even terms. The Taylor series expansion of this function contains a first-order term $k\Delta v$, which when multiplied by the proper time $s/\Delta v$ leaves a residual term, $sk$, that does not go to zero as $\Delta v$ goes to zero. Thus, in an inertial frame that is anisotropic, the slow separation of synchronized clocks does not preserve the synchronization of the clocks.

Consider two inertial frames K and K′ that have a relative velocity of u. The isotropic time dilation function of K, which is an even function of $\Delta v$, becomes anisotropic when extended to K′. When observers in frame K observe a clock moving at $\Delta v$ with respect to K′, the additional time dilation experienced by this clock with respect to clocks stationary in K′ depends on whether $\Delta v$ is along the direction of relative motion between the two frames or opposite to it. In the former case, the additional time dilation is a function of $(u + \Delta v)$, and in the latter case it is a function of $(u - \Delta v)$. Although an even function of $\Delta v$ satisfies the identity $f(\Delta v) = f(-\Delta v)$, it does not satisfy the relation $f(u + \Delta v) = f(u - \Delta v)$. Thus, observers in K will find the slow separation of clocks in K′ to be anisotropic and therefore will not expect a correct synchronization of these clocks. Observers in K will consider the synchronization obtained in K′ by the procedure of slow separation of clocks to be incorrect. Observers in K′, to whom their frame is isotropic, will however, consider the synchronization thus obtained to be correct. A similar account will be given by observers in K′ about the clock synchronization by slow separation of clocks in K, with u replaced by –u. Thus, the relative motion between K and K′ produces different synchronizations for K and K′ by the method of slow separation of clocks. If an observer at rest in an inertial frame K synchronizes two or more clocks at rest in K by the procedure of slow separation of clocks, an observer at rest in an inertial frame K′ that is moving with respect to K, will not consider those clocks at rest in K to be correctly synchronized. The same difference is reproduced when a light ray is used by frames K and K′ with the assumption that the speed of light is constant and isotropic in their respective frames. For observers at rest in K, both these procedures are successful only in an isotropic inertial frame and both the procedures fail in K′ because K′ is anisotropic. The fact that both these procedures produced identical synchronizations does not necessarily imply that the synchronization thus achieved is correct because, for observers at rest in K, both these procedures are in error by the same amount in an anisotropic inertial frame.[11]

Thus, the statement that "in an isotropic inertial frame, when identical and synchronized clocks present initially at the same location are separated very slowly, the synchronization of these clocks is almost unaffected" leads to two other results that can be stated as follows: (1) In an anisotropic inertial frame the slow separation of clocks does not necessarily produce a set of spatially separated and synchronized clocks. (2) When an inertial frame utilizes the process of slow separation of clocks to define a set of spatially separated and synchronized clocks, that inertial frame will be isotropic to an observer at rest in that inertial frame, under the synchronization thus obtained.

### III. ISOTROPY AND SYNCHRONIZATION

The results in Sec. II elucidate the circular relation between isotropy and synchronization of clocks. We also note that in the context of special relativity and a given set of spatially separated and synchronized clocks in an inertial reference frame, any other inertial reference frame moving with respect to this frame is anisotropic. However, an observer in any inertial frame can define synchronization in such a way as to render that frame isotropic.

The method of synchronizing with a light ray is easier to implement than the slow separation of clocks, especially over large distances. In an isotropic inertial frame the two methods produce identical and correct synchronization, and in an anisotropic inertial frame the two methods produce identical but incorrect synchronization. However, two inertial frames K and K′ that are in relative motion disagree with each other's synchronization and both are intrinsically isotropic; that is, observers at rest in K consider K to be isotropic and observers at rest in K′ consider K′ to be isotropic. It is believed that synchronization must be universal,[16, 17, 18] in the sense that the synchronization obtained in all inertial frames should be identical. This school of thought assumes that the synchronization of spatially separated clocks obtained in an isotropic inertial frame by the Einstein synchronization procedure is valid for other inertial frames as well.[13, 14, 16-18] If the synchronization obtained in other inertial frames by the Einstein synchronization procedure does not agree with this synchronization, it is because the other inertial frames are anisotropic, and the Einstein synchronization procedure fails to produce the correct synchronization in anisotropic inertial frames. Another school of thought believes that synchronization is a self-contained property within an inertial frame only and that different inertial frames can have different synchronizations of spatially separated clocks.[2, 5, 9, 19] The notion of universal synchronization, valid across all inertial frames, apparently assumes that there exists a unique stationary, isotropic inertial frame and other inertial frames are in error in neglecting their uniform motion and assuming themselves to be isotropic. The relativistic view is that all inertial frames are equivalent[9, 19] and each one of them can assume itself to be stationary and isotropic.

The conventional argument can be summarized as follows: when one considers two inertial frames K and K′ that are in relative motion with respect to each other, both K and K′ cannot be truly isotropic. In a truly isotropic frame the Einstein synchronization procedure or the procedure of slow separation of clocks produces the correct synchronization. However, an anisotropic inertial frame is observed to be isotropic by observers at rest in that frame, when they assume that the synchronization obtained by either of these procedures is correct. This assumption is valid only in an isotropic inertial frame. Thus the observed isotropy of an anisotropic inertial frame can be due to inaccurate synchronization of clocks. In contrast, the school of thought that believes in the "equivalence of inertial frames" subscribes to the argument that inertial frames K and K′ are equivalent and both are isotropic.

In attempting to understand the clock synchronization problem, students of special relativity often need to answer the question of whether moving clocks actually run slow or appear to run slow. If the answer to this question is "yes, they do run slow," then the principle of "equivalence of inertial frames" is apparently violated. If the answer is "no, they do not run slow, but only appear to run slow," then the question arises as to why moving clocks appear to run slow. This latter question is resolved in special relativity by proposing an integrated 4-dimensional space-time continuum.[9, 19] Therefore, according to special relativity, uniform motion is actually a "rotation" in the space-time continuum. The concept of the 4-dimensional space-time continuum maintains the equivalence of the inertial frames that have contracted objects and slow moving clocks as observed by each other. The conventionalists,[16, 17] however, advocate universal synchronization, a unique preferred isotropic frame, actual contraction of moving objects, and slow running of moving clocks. The conventionalists maintain that the re-synchronization or change in the preferred synchronization in the moving frame is only a convenience that results in apparent isotropy in the moving frame. Janis[10] concludes "The debate about conventionality of simultaneity seems far from settled, although some proponents on both sides of the argument might disagree with that statement." It is also recognized that no experiment can possibly be designed to resolve the issue because of the circular relation between clock synchronization and the measurement of the one-way speed of light.[2]

The adoption of different synchronizations by different inertial frames results in apparent causal anomalies. We briefly describe two of these[20, 21] to illustrate how the time order of events and the measurements of lengths and angles are affected by a dynamic synchronization convention. In the collision of inclined rods[20] the time-order of the collisions at the two ends of the rods is observed to be reversed by the two inertial frames co-moving with each of the rods. In the resolution of the rod-slot paradox[21] it has been shown that whether the rod is shorter or longer than the slot and the inclination at which the rod enters the slot depends on the synchronization convention associated with the inertial frames co-moving with the rod and the slot.

The disagreement concerning the synchronization of spatially separated clocks between observers of different inertial frames is known as the relativity of simultaneity. The circular relations between the one-way speed of light, isotropy, and clock synchronization do not permit an objective measurement of the one-way speed of light.[2] What is experimentally verifiable is that the average round-trip speed of light is constant in all inertial frames.

The main contribution of this paper is to show that the principle of relativity of simultaneity is not essential to satisfy the requirement that the average round-trip speed of light is constant in all inertial frames. We derive an expression for the one-way speed of light in an inertial frame K′ moving with respect to another inertial frame K, while maintaining the Einstein synchronization of K as the synchronization of K′ as well. This expression maintains the constancy of the average round-trip speed of light. We extend this result by proposing another expression, Eq. (39), for the one-way speed of light that includes a parameter. The value of this parameter depends on which inertial frame's Einstein synchronization is extended to all inertial frames. This expression also maintains the average round-trip speed of light to be constant.

**IV. FORMULATION OF ONE-WAY SPEED OF LIGHT**

The Lorentz transformation is the mathematical formulation[9] of the transformation of event coordinates between two inertial frames with relative velocity $\vec{v}$ between them. This transformation is symmetric in that the inverse transformation is obtained by replacing $\vec{v}$ with $-\vec{v}$ and interchanging the coordinates and times of the two frames.

Let K′ be an inertial frame observed by another inertial frame K to be moving with velocity $\vec{v}$ along the positive x-direction of K, which is also the positive x′-direction of K′. The spatial origins of the two frames coincide at t = t′ = 0. Define γ as

$$\gamma = \frac{1}{\sqrt{1 - v^2/c^2}}, \tag{1}$$

where $c$ is the average round-trip speed of electromagnetic radiation in vacuum in all inertial frames. When the Einstein synchronization procedure is adopted by an inertial frame, $c$ becomes the isotropic speed of electromagnetic radiation in vacuum in that inertial frame.

The inertial frame K observes the following about the inertial frame K′:

(1) movement of all objects at rest in K′ at velocity $\vec{v}$ along the line of motion;

(2) Lorentz-Fitzgerald contraction of rulers at rest in K′ along the line of motion by a factor of $1/\gamma$ as observed in K;

(3) slow running of clocks at rest in K′ by a factor of $1/\gamma$ as observed by clocks of K, or equivalently, clocks in K′ to an observer in K run slower by a factor of $1/\gamma$;

(4) clocks in K′ are asynchronous, assuming that the observers at rest in K and K′ have each synchronized the clocks at rest in their own frames using the Einstein synchronization procedure.

In accordance with the Lorentz transformations, the asynchronous observation (4) is well established such that at a given instant in K, clocks of K′ (that are synchronous as observed by K′) are asynchronized by $vx\gamma/c^2$ or $vx'/c^2$ as observed by K, where x and x′ are the distance of separation between the two spatial locations as observed by K and K′ respectively. This asynchronization is a necessary component for rendering the forward and inverse Lorentz transformations identical except for the change in the sign of v.

The four observations can be characterized by the following matrices in one spatial dimension, which act on the event coordinates (x, t) of K to produce the event coordinates (x′, t′) of K′:

$$\begin{pmatrix} 1 & -v \\ 0 & 1 \end{pmatrix}, \begin{pmatrix} \gamma & 0 \\ 0 & 1 \end{pmatrix}, \begin{pmatrix} 1 & 0 \\ 0 & 1/\gamma \end{pmatrix}, \begin{pmatrix} 1 & 0 \\ -v/c^2 & 1 \end{pmatrix}. \tag{2}$$

As a result, observers in frame K find the composition of the transformation of event coordinates from frame K to frame K′ (Lorentz transformation) to be

$$\begin{pmatrix} 1 & 0 \\ -v/c^2 & 1 \end{pmatrix} \begin{pmatrix} 1 & 0 \\ 0 & 1/\gamma \end{pmatrix} \begin{pmatrix} \gamma & 0 \\ 0 & 1 \end{pmatrix} \begin{pmatrix} 1 & -v \\ 0 & 1 \end{pmatrix} = \begin{pmatrix} \gamma & -v\gamma \\ -v\gamma/c^2 & \gamma \end{pmatrix}. \qquad (3)$$

Because the second and third matrices on the left-hand side of Eq. (3), representing the effects of length contraction and time dilation, respectively, are diagonal, they commute. Therefore, these two matrices can be combined without loss of generality and Eq. (3) can be rewritten as:

$$\begin{pmatrix} 1 & 0 \\ -v/c^2 & 1 \end{pmatrix} \begin{pmatrix} \gamma & 0 \\ 0 & 1/\gamma \end{pmatrix} \begin{pmatrix} 1 & -v \\ 0 & 1 \end{pmatrix} = \begin{pmatrix} \gamma & -v\gamma \\ -v\gamma/c^2 & \gamma \end{pmatrix}. \qquad (4)$$

The right-hand side is the Lorentz transformation of event coordinates between two inertial frames in relative motion. We will utilize Eq. (4) to derive a constructive formulation for the one-way speed of light.

We now examine each component of the decomposition and study the transformed speed of light after each step. We begin by rewriting the matrices on the left-hand side of Eq. (4) in four dimensions – three spatial and one temporal. We use the order of the variables [x,t,y,z] so that the extension from one to three spatial dimensions is straightforward. The *x* axis of inertial frame K and the *x'* axis of inertial frame K' are chosen such that the line of relative motion coincides with the *x* and *x'* axes.

(1) The Galilean transformation is given by

$$\begin{pmatrix} 1 & -v & 0 & 0 \\ 0 & 1 & 0 & 0 \\ 0 & 0 & 1 & 0 \\ 0 & 0 & 0 & 1 \end{pmatrix} = A, \qquad (5)$$

which transforms the event coordinates [ *x, t, y, z*] of an event as measured by observers at rest in inertial frame K, to the event coordinates [*x', t', y', z'*] of the same event as measured by observers at rest in inertial frame K'.

(2) The Lorentz-Fitzgerald length contraction and time dilation is given by

$$\begin{pmatrix} \gamma & 0 & 0 & 0 \\ 0 & 1/\gamma & 0 & 0 \\ 0 & 0 & 1 & 0 \\ 0 & 0 & 0 & 1 \end{pmatrix} = B. \qquad (6)$$

The transformation in Eq. (6) refers to Lorentz-Fitzgerald length contraction and time dilation. The rulers of the moving frame are contracted by a factor $\gamma$ and therefore a given length along the *x* direction is measured to be increased by a factor γ by the moving frame; any particular clock of the moving frame, when observed by clocks of the stationary frame, runs slower by a factor γ.

(3) A specific re-synchronization as applicable to K' is given by a transformation that describes the relative asynchronization of clocks between the two frames:

$$\begin{pmatrix} 1 & 0 & 0 & 0 \\ -v/c^2 & 1 & 0 & 0 \\ 0 & 0 & 1 & 0 \\ 0 & 0 & 0 & 1 \end{pmatrix} = D.  \tag{7}$$

The Lorentz transformation in three-dimensional space is obtained as the matrix product DBA.

$$DBA = \begin{pmatrix} \gamma & -v\gamma & 0 & 0 \\ -v\gamma/c^2 & \gamma & 0 & 0 \\ 0 & 0 & 1 & 0 \\ 0 & 0 & 0 & 1 \end{pmatrix}. \tag{8}$$

Note that DBA = A B$^{-1}$D are two forms of the decomposition of the Lorentz transformation that are a consequence of the non-commutative nature of the superposition of transformations.[22]

Consider a light ray emitted from the origin of frame K at $t = 0$ and speed $c$ in a given direction that makes an angle $\theta$ with the *z*-axis. Let the projection of that direction on the *x-y* plane make an angle $\phi$ with the positive *x*-axis. At some instant *t* in frame K, the following event is generated:

$$[x, t, y, z] = [ct\sin\theta\cos\phi, t, ct\sin\theta\sin\phi, ct\cos\theta], \tag{9}$$

where $\theta$ and $\phi$ are the usual spherical coordinates, with $0 \leq \theta \leq \pi$ and $0 \leq \phi < 2\pi$. The transformation denoted by matrix A in Eq. (5) transforms the event coordinates given in Eq. (9) to:

$$[x, t, y, z] = [(c\sin\theta\cos\phi - v)t, t, ct\sin\theta\sin\phi, ct\cos\theta]. \tag{10}$$

The resultant velocity is

$$\vec{c}_g = [c\sin\theta\cos\phi - v, c\sin\theta\sin\phi, c\cos\theta], \tag{11}$$

where $\vec{c}_g$ denotes the velocity of light as observed in a reference frame moving at speed v along the x direction with respect to an inertial frame in which light propagates isotropically at speed $c$. Equation (11) is valid only under Newtonian mechanics when the time coordinate is invariant across all inertial frames and there is no observed length contraction of moving objects. In this case the resultant speed is not equal to $c$ and neither is the average round-trip speed.

The event coordinates in Eq. (10) are transformed by the matrix B given in Eq. (6) to:

$$[x, t, y, z] = \left[(c\sin\theta\cos\phi - v)\gamma t, \frac{t}{\gamma}, ct\sin\theta\sin\phi, ct\cos\theta\right]. \quad (12)$$

We now show from the event coordinates in Eq. (12) that the average round-trip speed of light in K' is a constant and equal to $c$, but the speed is different in different directions. We denote

$$t' = \frac{t}{\gamma}, \quad (13)$$

and rewrite Eq. (12) as

$$[x, t, y, z] = [(c\sin\theta\cos\phi - v)\gamma^2 t', t', c\gamma t'\sin\theta\sin\phi, c\gamma t'\cos\theta]. \quad (14)$$

The velocity in K' is

$$\vec{c}' = [(c\sin\theta\cos\phi - v)\gamma^2, c\gamma\sin\theta\sin\phi, c\gamma\cos\theta]. \quad (15)$$

The resultant speed in K' is

$$c' = c\gamma^2\left(1 - \frac{v}{c}\sin\theta\cos\phi\right). \quad (16)$$

We divide each of the components in Eq. (15) by the right-hand side of Eq. (16) and obtain the direction cosines of the propagation as observed in the transformed co-ordinates:

$$\sin\theta'\cos\phi' = \frac{(\sin\theta\cos\phi - v/c)}{1 - \frac{v}{c}\sin\theta\cos\phi}, \quad (17)$$

$$\sin\theta'\sin\phi' = \frac{\sin\theta\sin\phi}{\gamma\left(1 - \frac{v}{c}\sin\theta\cos\phi\right)}, \quad (18)$$

and

$$\cos\theta' = \frac{\cos\theta}{\gamma\left(1 - \frac{v}{c}\sin\theta\cos\phi\right)}. \quad (19)$$

Given $\theta$ and $\phi$, $\theta'$ can be evaluated from Eq. (19) and $\phi'$ can be evaluated from either Eq. (17) or Eq. (18).

If we consider the plane formed by the line of relative motion between the inertial frames K and K' and the line of the light ray's travel, and take this plane to be the *x-y* plane, we have $z = 0$, $z' = 0$, $\theta = \pi/2$, and $\theta' = \pi/2$. By substituting these values into Eqs. (17) and (18) and dividing appropriately, we obtain the standard aberration formula,

$$\tan\phi' = \frac{\sin\phi}{\gamma(\cos\phi - \frac{v}{c})}. \tag{20}$$

An important difference is that although the standard method of deriving the aberration formula utilizes the Lorentz transformation DBA, we have derived it using the transformation BA, which is the Lorentz transformation devoid of the synchronization shift represented by matrix D. The transformation BA is known as the inertial-synchronized-Tangherlini transformation.[17, 23, 24] Thus the aberration results from the Lorentz transformation and the inertial-synchronized-Tangherlini transformation are identical because the latter transformation (BA) determines the direction of propagation in K'. The additional component of the Lorentz transformation, represented by matrix D, does not change this direction of propagation, but adjusts only the clock synchronization to make the speed of light equal to *c*, in accordance with the Einstein synchronization convention.

From Eq. (17) we obtain

$$\sin\theta\cos\phi = \frac{\sin\theta'\cos\phi' + v/c}{1 + \frac{v}{c}\sin\theta'\cos\phi'}. \tag{21}$$

By substituting Eq. (21) into Eq. (16) we obtain the speed of propagation in the transformed coordinates to be

$$c'_{out} = \frac{c}{1 + \frac{v}{c}\sin\theta'\cos\phi'} = \frac{c}{1 + \hat{\mathbf{c}}'\cdot\frac{\mathbf{v}}{c}}, \tag{22}$$

where $\hat{\mathbf{c}}'$ is a unit vector in the direction of the outgoing light wave as observed in K' and **v** is the velocity of K' relative to K. For the reverse direction with angles $\pi - \phi'$ and $\pi - \theta'$, we obtain the speed of propagation as:

$$c'_{ret} = \frac{c}{1 - \frac{v}{c}\sin\theta'\cos\phi'} = \frac{c}{1 - \hat{\mathbf{c}}'\cdot\frac{\mathbf{v}}{c}}. \tag{23}$$

The average round-trip speed of the light ray is given by the total distance (outward + return) divided by the total time (outward journey time + return journey time). Because the outward distance and the return distance are the same, the average round-trip speed becomes the harmonic mean of the outward and return speeds, namely $2/(1/c'_{out} + 1/c'_{ret})$. It is easily verified that the harmonic mean of the speeds given in Eqs. (22) and (23) is equal to $c$.

From Eq. (22) we can deduce the Reichenbach parameter $\varepsilon$ in three-dimensional Euclidean space to be

$$\varepsilon = \frac{1}{2}(1 + \frac{v}{c}\sin\theta'\cos\phi'), \tag{24}$$

using the relation $c' = c/2\varepsilon$ (as explained in Ref. 2). Equation (24) defines the directional dependence of $\varepsilon$. We can also infer that there exists a unique direction along which $\varepsilon$ is maximum ($\varepsilon_{max}$). Once the unique direction and $\varepsilon_{max}$ are known, the value of $\varepsilon$ in any other direction is determined from Eq. (24), where $v = c(2\varepsilon_{max} - 1)$. The unique direction along which $\varepsilon$ is maximum is the direction of relative motion of the inertial frame with respect to an isotropic inertial frame. This direction is also the chosen x'-axis and $\theta'$ and $\phi'$ are measured accordingly. This analysis clarifies the relation between $\varepsilon$ and the implicit velocity of the inertial frame with respect to the isotropic frame. It also emphasizes the directional dependence of $\varepsilon$.

If we apply the last component D of the matrix composition to the event coordinates as given in Eq. (14), we obtain:

$$\begin{pmatrix} 1 & 0 & 0 & 0 \\ -v/c^2 & 1 & 0 & 0 \\ 0 & 0 & 1 & 0 \\ 0 & 0 & 0 & 1 \end{pmatrix} \begin{pmatrix} (c\sin\theta\cos\phi - v)\gamma^2 t' \\ t' \\ c\gamma t'\sin\theta\sin\phi \\ c\gamma t'\cos\theta \end{pmatrix}. \tag{25}$$

By appropriate substitution from Eqs. (17)–(19) and Eq. (21), Eq. (25) can be shown to be equal to

$$\begin{pmatrix} ct''\sin\theta'\cos\phi' \\ t'' \\ ct''\sin\theta'\sin\phi' \\ ct''\cos\theta' \end{pmatrix}, \tag{26}$$

where

$$t'' = t'\left[1 - \frac{v}{c^2}\gamma^2(c\sin\theta\cos\phi - v)\right]. \tag{27}$$

Because $t' = t/\gamma$, $t''$ can be expressed as

$$t'' = (\frac{t}{\gamma} + \frac{v^2 t\gamma}{c^2}) - \frac{v\gamma}{c^2} ct\sin\theta\cos\phi, \qquad (28)$$

which can be simplified to

$$t'' = t\gamma - \frac{v\gamma}{c^2} ct\sin\theta\cos\phi. \qquad (29)$$

If we use the coordinates of the event specified in Eq. (9), $x = ct\sin\theta\cos\phi$, we obtain

$$t'' = (t - \frac{vx}{c^2})\gamma, \qquad (30)$$

which is in accordance with the Lorentz transformation.

Both $t'$ and $t''$ are bona fide time coordinates for specifying the time of occurrence of an event in frame K′. The physical differences between them are summarized in Table I.

|   | Physical observation | K′ uses $t'$ as time coordinate of events | K′ uses $t''$ as time coordinate of events |
|---|---|---|---|
| 1 | Agreement with observers in K on synchronization of spatially separated clocks | Agree | Disagree |
| 2 | Speed of light in any direction | Variable | Constant |
| 3 | Average round-trip speed of light in any closed path | Constant | Constant |
| 4 | Transformation of event coordinates from K to K′ | Matrix product BA (= IST transformation) | Matrix product DBA = AB$^{-1}$D (= Lorentz transformation) |

**Table I. Comparison of observations for the IST and Lorentz transformations**

The IST transformation[16] implies that observers in frame K′ retain the clock synchronization as obtained in frame K by the Einstein synchronization procedure as applicable to K. The coordinates given in Eq. (26) indicate that the ray of light originally observed to be propagating at $c$ in the direction $(\theta, \phi)$ is still observed to be propagating at $c,$ but in the direction $(\theta', \phi')$. The coordinates in Eq. (26) are obtained from the coordinates of Eq. (9) by the Lorentz transformation as applicable to spherical coordinates. The relationship between the primed and unprimed cosines is given in Eqs. (17)–(19).

Therefore, although the transformation BA maintains the average two-way speed of light to be a constant, the transformation DBA, which is the Lorentz transformation, maintains not only the average two-way speed of light, but also the one-way speed of light to be a constant.

## V. DISCUSSION

We give here a generalization of the development in Sec. IV. Suppose we have a linear transformation of event coordinates in one-dimensional space as:

$$\begin{pmatrix} a_{11} & a_{12} \\ a_{21} & a_{22} \end{pmatrix} \begin{pmatrix} x \\ t \end{pmatrix} = \begin{pmatrix} x' \\ t' \end{pmatrix}. \tag{31}$$

Let K denote the inertial frame associated with coordinates $(x, t)$ and K′ be the inertial frame associated with coordinates $(x', t')$. If there is a uniform relative motion between K and K′, then $a_{12} = -v\, a_{11}$, where v is the speed of K′ relative to K.

If we consider a light ray observed by K and K′ to originate from the origin of K′ and return to the origin of K′ after reflection by a mirror, we can describe the observations of frame K as follows:

Event $E_1$: Leaving of light ray from origin of K and K′, which are coincident at this event.
Event $E_2$: Light ray reaching a mirror located at a distance $ct_1$ at time $t_1$.
Event $E_3$: Light ray after reflection returns to position $c(t_1 - t_2)$ at time $(t_1 + t_2)$.

Table II gives the transformed coordinates in frame K′.

|  | **Frame K** | **Frame K′** |
|---|---|---|
| $E_1$ | *(0, 0)* | *(0, 0)* |
| $E_2$ | *($ct_1$, $t_1$)* | [($a_{11}ct_1 + a_{12}t_1$), ($a_{21}ct_1 + a_{22}t_1$)] |
| $E_3$ | [*$c(t_1 - t_2)$, $(t_1 + t_2)$*] | [($a_{11}c(t_1 - t_2) + a_{12}(t_1 + t_2)$), ($a_{21}c(t_1 - t_2) + a_{22}(t_1 + t_2)$)] |

**Table II. Event coordinates of key events on the path of a light ray originating from and returning to the origin of frame K′**

The light ray returns to the origin of frame K′ at $E_3$. Therefore, the spatial coordinate at $E_3$ in frame K′ is zero, which implies that

$$a_{11}c(t_1 - t_2) + a_{12}(t_1 + t_2) = 0, \tag{32}$$

or

$$(t_1 - t_2)/(t_1 + t_2) = -a_{12}/(a_{11}c) = (v/c) \tag{33}$$

If we assume the average round-trip speed of light to be $c$, we have

$$c = 2(a_{11}ct_1 + a_{12}t_1)/(a_{21}c(t_1 - t_2) + a_{22}(t_1 + t_2)). \tag{34}$$

We take into account $(v/c) = (t_1 - t_2) / (t_1 + t_2)$ and $a_{12} = -v a_{11}$ and simplify Eq. (34) to

$$a_{22} = \frac{a_{11}}{\gamma^2} - a_{21} v. \tag{35}$$

If we impose the condition that the round-trip average speed of light is to be equal to $c$, we obtain

$$\begin{pmatrix} a_{11} & -a_{11} v \\ a_{21} & (\frac{a_{11}}{\gamma^2} - a_{21} v) \end{pmatrix} \begin{pmatrix} x \\ t \end{pmatrix} = \begin{pmatrix} x' \\ t' \end{pmatrix}. \tag{36}$$

We further impose the condition that the determinant of this transformation be unity, that is, the scale factor of the transformation is unity, which ensures that the property that the average round-trip speed of light is $c$, can be generalized to two- and three-dimensional spaces. When we set the determinant of the matrix in Eq. (36) to unity, we obtain

$$a_{11} = \gamma = \frac{1}{\sqrt{1 - v^2/c^2}}. \tag{37}$$

We set $a_{11} = \gamma$ and include the two other invariant spatial dimensions that are perpendicular to the line of relative motion so that the transformation becomes

$$\begin{pmatrix} \gamma & -v\gamma & 0 & 0 \\ a_{21} & \left(\frac{1}{\gamma} - a_{21} v\right) & 0 & 0 \\ 0 & 0 & 1 & 0 \\ 0 & 0 & 0 & 1 \end{pmatrix} \begin{pmatrix} x \\ t \\ y \\ z \end{pmatrix} = \begin{pmatrix} x' \\ t' \\ y' \\ z' \end{pmatrix}, \tag{38}$$

where $a_{21}$ is a parameter. Equation (38) transforms the event coordinates [$x$, $t$, $y$, $z$] of an event as observed in an isotropic inertial frame K to the event coordinates [$x'$, $t'$, $y'$, $z'$ ] of the same event as observed in another inertial frame K' that is in relative motion with respect to K. The line of relative motion coincides with the $x$ and $x'$ axes. The transformation maintains the average round-trip speed of light in any closed path in K' to be a constant equal to the isotropic light propagation speed in frame K.

By combining the result in Eq. (38) with the development in Sec. IV, we can derive the following generalized one-way speed of light in three-dimensional Euclidean space to be

$$c'_{out} = \frac{c}{1 + \left(\frac{v}{c} + \frac{a_{21} c}{\gamma}\right) \sin\theta' \cos\phi'} = \frac{c}{1 + \hat{\mathbf{c}}' \bullet \frac{\mathbf{v}}{c}\left(1 + \frac{c^2}{v\gamma} a_{21}\right)}. \tag{39}$$

If we substitute $a_{21} = 0$ (for the IST transformation), the speed is $\dfrac{c}{1 + \dfrac{v}{c}\sin\theta'\cos\phi'}$, as given by Eq. (22). The IST transformation[16] implies that observers in frame K′ retain the clock synchronization as obtained in frame K by the Einstein synchronization procedure as applicable to K. For $a_{21} = -v\gamma/c^2$ (the Lorentz transformation) the speed is equal to $c$. It can also be verified from Eq. (39) that for any value of $a_{21}$, the average round-trip speed is equal to $c$. Other values of $a_{21}$ imply that frame K′ has adopted the Einstein synchronization convention of another inertial frame that is in uniform motion with respect to both K and K′.

The allowed range of $a_{21}$ can be determined as follows. Because the speed of light must be positive and because $\theta'$ and $\phi'$ are independent variables, we obtain $\left|\dfrac{v}{c} + a_{21}\dfrac{c}{\gamma}\right| < 1$. This condition defines the permissible range of values of $a_{21}$ for a given value of v. Because $\gamma$ and $c$ are positive, the upper limit of the range is $(\gamma/c)(1 - (v/c))$, which can be simplified to $\dfrac{1}{c}\sqrt{\dfrac{1 - v/c}{1 + v/c}}$. Similarly, the lower limit is $(-\gamma/c)(1 + (v/c))$, which simplifies to $\dfrac{-1}{c}\sqrt{\dfrac{1 + v/c}{1 - v/c}}$.

The value of the upper limit goes to zero as v approaches $+c$. The value of the lower limit goes to zero as v approaches $-c$. Thus, the minimum value of the upper limit and the maximum value of the lower limit are both zero. The values of $a_{21} = 0$ and $a_{21} = -v\gamma/c^2$ both fall within this range when $|v/c| < 1$.

If we assume $a_{21}$ to be a constant independent of v, then the only possible value for $a_{21}$ is zero. This assumption leads to the IST transformation. If we assume $a_{21}$ to be a function of v, then the choice which lies at the center of the permissible range, $a_{21} = \dfrac{-v\gamma}{c^2}$, leads to the Lorentz transformation.

## VI. CONCLUSIONS

We have considered four observations by an observer in frame K of events in frame K′. These are relative movement, contraction of rulers, slow running of clocks, and asynchronization of the clocks of K and K′. Of these four observations, the first three, when superimposed one by one in a specified order, satisfy the requirements of the Michelson-Morley experiment. The inclusion of the fourth observation, the asynchronization between inertial frames as denoted by matrix D of Eq. (7), satisfies an additional requirement imposed by the two postulates of special relativity that all inertial frames are isotropic and equivalent. The inclusion of the asynchronization component leads to the possibility of an isotropic one-way speed of light in frame K′, but this asynchronization is not implied by the Michelson-Morley experiment.

The adoption of a non-Einstein synchronization by an inertial frame implies that the inertial frame assumes another inertial frame to be isotropic and consequently considers itself to be anisotropic. This anisotropy is due to its uniform motion with respect to an isotropic inertial frame. The value of ε in different directions is dictated by Eq. (24), and the direction of motion is determined by the direction in which ε is maximum. An inertial frame adopting a synchronization other than Einstein synchronization will have to specify a direction in which ε is maximum. Thus, it implicitly recognizes its motion along that direction with respect to an isotropic inertial frame.

In the context of dynamics, Ohanian's arguments[2] imply that any choice of the parameter $a_{21}$ other than the choice that leads to the Lorentz transformation does not preserve the form of Newton's second law of motion. However, given a universal synchronization convention, Lorentz-Fitzgerald length contraction, and time dilation, only one inertial frame is isotropic. Newton's laws are obeyed in the unique isotropic frame only when the velocities are small ($v^2 \ll c^2$). In other inertial frames, Newton's laws are valid (only for small velocities) when the Einstein synchronization is adopted in that inertial frame. However, Einstein synchronization is the correct synchronization only in the isotropic inertial frame. An anisotropic inertial frame that adopts the Einstein synchronization convention, as applicable to itself, presupposes that it is isotropic.

In three-dimensional Euclidean space the IST and Lorentz transformations are special cases of a more general transformation that has a parameter as specified in Eq. (38). It transforms the event coordinates of an event in an isotropic inertial frame K to the event coordinates of the same event in another inertial frame K′ that is anisotropic. For any value of the parameter $a_{21}$ within the permissible range discussed in Sec. V, the inertial frame K′, even though anisotropic, maintains the property that the average round-trip speed of light is constant. A particular choice of the parameter $a_{21}$ corresponds to the choice of a corresponding particular inertial frame whose Einstein synchronization is adopted as the synchronization in the inertial frame K′ as well. If the parameter $a_{21}$ is chosen to be zero, it implies that the Einstein synchronization of frame K is also the synchronization followed by K′. In this case, the transformation becomes the IST transformation. When the parameter $a_{21}$ is chosen to be $-v\gamma/c^2$, it implies that observers in frame K′ adopt the Einstein synchronization as applicable to K′ to measure time coordinates of events; only for this specific choice of $a_{21}$, does K′ become intrinsically isotropic and the transformation specified by Eq. (38) becomes the Lorentz transformation. If there exists a unique isotropic inertial frame, then any inertial frame that is in uniform motion with respect to the isotropic inertial frame can continue to use the Einstein synchronization of the isotropic inertial frame. Thus the existence of a unique isotropic inertial frame is compatible with the average round-trip speed of light being constant in all inertial frames.

**ACKNOWLEDGMENTS**

The authors are indebted to the referees for greatly improving the presentation of this paper. We are particularly grateful to one of the referees for many valuable suggestions, especially in Secs. I–III . We would also like to thank two colleagues for their helpful suggestions in improving the readability of the paper.

**REFERENCES**

[1] A. Michelson and E. Morley, "On the relative motion of the earth and the luminiferous Ether," Am. J. Sci. **34** (203), 333-345 (1887).

[2] H. Ohanian, "The role of dynamics in the synchronization problem," Am. J. Phys. **72** (2), 141-148 (2004).

[3] A. Macdonald, "Comment on 'The role of dynamics in the synchronization problem,'" Am. J. Phys. **73** (5), 454-455 (2005).

[4] A. Martinez, "Conventions and inertial reference frames," Am. J. Phys. **73** (5), 452-454 (2005).

[5] H. Ohanian, "Reply to Comment (s) on 'The role of dynamics in the synchronization problem by A. Macdonald, [Am. J. Phys. **73**, 454 (2005)] and A. Martinez [Am. J. Phys. **73**, 452 (2005)," Am. J. Phys. **73** (5), 456-457 (2005).

[6] A. S. Eddington, *The Mathematical Theory of Relativity* (Cambridge University Press, Cambridge, 1923).

[7] H. Reichenbach, *The Philosophy of Space and Time* (Dover, New York, 1957). First published in German under the title *Philosophie der Raum-Zeit-Lehre* in Walter de Gruyter, Berlin 1927.

[8] See particularly the references cited in the first two footnotes of Ref. 2.

[9] A. Einstein, in *Relativity, The Special and General Theory* (Three Rivers Press, New York, 1961).

[10] A. Janis, "Conventionality of simultaneity," in The Stanford Encyclopedia of Philosophy, edited by Edward N. Zalta, Published Online. Location: Stanford Year 1998 Revised 2006

http://plato.stanford.edu/entries/spacetime-convensimul/

[11] The error in the method of slow separation of clocks is due to the directional dependence of the rate at which clocks run when moved at a specified speed in different directions in an anisotropic inertial frame. The error in synchronizing clocks by assuming

the speed of light is constant in all directions is due to the invalidity of this assumption in an anisotropic inertial frame. In an isotropic inertial frame both methods are accurate and produce identical results. In an anisotropic inertial frame both methods are inaccurate but produce identical results.

[12] D. Bohm, *The Special Theory of Relativity* (W. A. Benjamin, New York, 1965), Chap. 8

[13] H. A. Lorentz, "Simplified theory of electrical and optical phenomena in moving systems," Proc. Acad. Sci. Amsterdam **1**, 427-442 (1899).

[14] H. A. Lorentz, "Electromagnetic phenomena in a system moving with any velocity smaller than that of light," Proc. Acad. Sci. Amsterdam **6**, 809-831 (1904).

[15] Both Ref. 12 and Ref. 9 propose a time dilation function $\sqrt{1 - v^2/c^2}$ that does not contain any term in $|v|$. The former is based on the conventional view of synchronization and the latter on the relativistic view of synchronization. The analysis by the Lorentz theory of electrons (Ref. 12, Chap. VII) uses differential equations and all functions are expected to be differentiable throughout the domain which includes the point v = 0.

[16] F. R. Tangherlini, "On energy-momentum tensor of gravitational field," Nuovo Cimento Suppl. **20**, 1-19 (1961).

[17] R. de Abreu and V. Guerra, "The principle of relativity and the indeterminacy of special relativity," Eur. J. Phys. **29**, 33-52 (2008).

[18] J. S. Bell, "How to teach special relativity," Progress in Scientific Culture **1**, 67-80 (1976).

[19] H. Minkowski, "Die Grundgleichungen fur die elektromagnetischen Vorgange in bewegten Korpern," Nachrichten Ges. Wiss. Gottingen, 53–116 (1908), reprinted in *Math Annalen* (with the same title) **68**, 472-525 (1910).

[20] C. Iyer and G. Prabhu, "Reversal in the time order of interactive events: the collision of inclined rods," Eur. J. Phys. **27**, 819-824 (2006).

[21] C. Iyer and G. Prabhu, "Differing observations on the landing of the rod into the slot," Am. J. Phys. **74** (11), 998-1001 (2006).

[22] C. Iyer, "Comment on 'The principle of relativity and the indeterminancy of special relativity,'" Eur. J. Phys. **29**, L13-L17 (2008).

[23] V. Guerra and R. de Abreu, "The conceptualization of time and the constancy of the speed of light," Eur. J. Phys. **26**, S117-S123 (2005).

[24] V. Guerra and R. de Abreu, "On the consistency between the assumption of a special system of reference and special relativity," Found. Phys. **36** (12), 1826-1845 (2006).